\documentclass[twocolumn,aps,prl,showpacs]{revtex4}
\usepackage{graphicx}% Include figure files
\usepackage{bm}% bold math
\def\molsim#1#2#3{{ Mol. Simulat.} {\bf #1}, #2 (#3).}

\def\physicaa#1#2#3{{ Physica A} {\bf #1}, #2 (#3).}
\def\arpc#1#2#3{{ Ann. Rev. Phys. Chem.} {\bf #1}, #2 (#3).}
\def\prl#1#2#3{{ Phys. Rev. Lett.} {\bf #1}, #2 (#3).}

\def\pre#1#2#3{{  Phys. Rev. E.} {\bf #1}, #2 (#3).}

\def\jcp#1#2#3{{ J. Chem. Phys.} {\bf #1}, #2 (#3).} 
\def\jpc#1#2#3{{ J. Phys. Chem} {\bf #1}, #2 (#3).}

\def\cpl#1#2#3{{ Chem. Phys. Lett.} {\bf #1}, #2 (#3).}

\def\science#1#2#3{{ Science} {\bf #1}, #2 (#3).}
\def\nature#1#2#3{ { Nature} {\bf #1}, #2 (#3).}

\def\sp#1{ $S_{#1}(f)$}

\def\onebyf{$1/f^\alpha$}

\def\avg#1{\langle #1\rangle}

\def\be{\begin{equation}}
\def\ee{\end{equation}}

\begin{document}
\title{Spectral Signatures of the Diffusional Anomaly in Water}
\author{Anirban Mudi}
\affiliation{Department of Chemistry, Indian Institute of Technology-Delhi,
New Delhi: 110016, India.}
\author{Ramakrishna Ramaswamy}
\affiliation{School of Physical Sciences, Jawaharlal Nehru University, 
New Delhi: 110067, India.}
\author{Charusita  Chakravarty}
\email{charus@chemistry.iitd.ernet.in}
\affiliation{Department of Chemistry, Indian Institute of Technology-Delhi,
New Delhi: 110016, India.}

\begin{abstract}
Analysis of power spectrum profiles for various tagged
particle quantities in bulk SPC/E water  is used to demonstrate that
variations in mobility associated with the diffusional anomaly
are mirrored in the exponent of the \onebyf\ region.
Monitoring of \onebyf\ behaviour is shown to be  a simple and direct
method for linking phenomena on three distinctive length and time 
scales: the local molecular environment, hydrogen bond network reorganisations
and the diffusivity.  The results indicate that experimental
studies of supercooled water to probe the density dependence of $1/f^\alpha$ 
spectral features, or equivalent stretched exponential behaviour
in time-correlation functions, will be of interest.
\end{abstract}

\maketitle

Supercooled water, unlike most liquids, shows an increase
in molecular mobility 
with compression over certain ranges of temperature and pressure 
\cite{ff72,caa83,plsl,afwb}. Along an isotherm, this anomalous diffusional 
regime is bounded by densities corresponding
to the diffusivity minimum and maximum 
\cite{ed01,ms98,sss99,nsbs,sslss,hpss}. In this work,
the diffusional anomaly
is related to dynamical variations in the special structural features of
water: a strong preference for local tetrahedral order and
the presence of a fluctuating three-dimensional network of hydrogen
bonds. 

We use power spectral analysis to study  temporal fluctuations in various 
observables. The power spectrum of a time-dependent mechanical
quantity, $A(t)$, is defined as
\be
S(f) = {\left| \int_{t_{min}}^{t_{max}} \bigl(A(t)-\avg{A}\bigr)
e^{2{\pi}ift}{ dt}\right|}^2.
\ee
 where $\avg A$ is the corresponding average over the
system trajectory.  In the case of water, power spectra associated
with several quantities show  a distinctive \onebyf\
type dependence on the frequency $f$ \cite{nrc95,sor92,mrc03,mc04}. 
Such \onebyf\ or flicker noise  is a generic feature of  systems with multiple
time scales \cite{em} and, in this context,
originates   from hydrogen bond network rearrangements involving variable 
numbers of molecules. This study shows that the power spectrum
profile associated with various quantities sensitive to the
local molecular environment, particularly in the \onebyf\ regime,
carries useful information on spatiotemporal correlations
in the hydrogen bond network, including distinct signatures
of the diffusional anomaly.

We have used the extended simple point charge (SPC/E) intermolecular potential 
for water\cite{bgs87}  since the density, diffusional and compressibility 
anomalies of this model are well-characterised 
\cite{sss99,nsbs,sslss,hpss}. 
Molecular dynamics (MD) simulations of  bulk SPC/E water in the NVT ensemble
were performed using the the DL\_POLY software package \cite{at86,syr01}. 
Simulation details are summarised in refs.\cite{mrc03,mc04,comp}. 

Figure 1 shows the power spectra associated with different tagged particle
quantities for bulk SPC/E water at 230K and 0.9 g~cm$^{-3}$. 
The tagged molecule potential energy is the interaction
energy of a given molecule with all other molecules in the system. The
corresponding power spectrum, \sp{u}, shows a broad 
peak, centred around 500 cm$^{-1}$,  due to the high-frequency,
essentially single-molecule, librational modes.  The $S_u(f)$ curve shows
two regions of \onebyf -type behaviour: (i)
the high frequency 60-298 cm$^{-1}$ range with $\alpha_u'= 1.56\pm 0.02$ and 
(ii) the low frequency 1-40 cm$^{-1}$ region with $\alpha{_u}=1.06 \pm 0.02$.
The O-O stretch and O-O-O bending modes
for bulk SPC/E water are known to occur at 200 and 50 cm$^{-1}$ respectively
\cite{ph98}
and therefore the high frequency
$1/f^{\alpha'}$ regime   involves two or three-molecule
hydrogen-bond network rearrangements.  Since increasing delocalisation
of vibrational modes results in lowering of the frequencies,
the low frequency  $1/f^{\alpha}$ regime
must involve displacements of four to six molecules.  Crossover to  
Markovian or white
noise behaviour should occur below 1 cm$^{-1}$ \cite{mrc03}
 which is a frequency regime not studied here. 
Also shown in Figure 1 is  \sp{lib} obtained from
fluctuations in the librational kinetic energy, defined
as the difference between the  total  and the centre-of-mass kinetic energies
of a rigid molecule.
\sp{lib} shows no evidence of multiple time scale behaviour and 
the single prominent peak is again due to the librational modes.

The local structural order around a tagged oxygen atom  is gauged by
two  order parameters, $q_O$ and $q_H$.
The degree of tetrahedrality of the four nearest  hydrogen atoms 
surrounding a given oxygen atom $i$ is measured by:
\be
q_H = 1 -\frac{3}{8}\sum_{j=1}^3\sum_{k=j+1}^4 (\cos \psi_{jk}+ 1/3)^2
\ee
where $\psi_{jk}$ is the angle between the bond vectors {\bf r}$_{ij}$
and {\bf r}$_{ik}$ where $j$ and $k$ label the four nearest hydrogen atoms.
The $q_O$ order parameter is similarly defined using the positions
of the four nearest oxygen atoms  and
has been previously used to characterise orientational order in water
\cite{ed01}.  The corresponding power spectra are  
labelled  \sp{O} and \sp{H}. 
The \sp{H} curve shows a clear peak due to librational motion, unlike
the \sp{O} curve.  Both \sp{O} and \sp{H} spectra  show two 
 \onebyf\  regimes with different exponents but in the case of \sp{O},
the higher frequency \onebyf\ regime  stretches between 200-1000 cm$^{-1}$.
The spectrum of fluctuations in the number of
nearest oxygen atoms surrounding a tagged oxygen atom, 
\sp{NN},  is very similar  to \sp{O}
\cite{nsbs}.

The key  features of the power spectral profiles
are: (i) a broad librational peak
(ii) a high-frequency $1/f^{\alpha'}$  regime above 50 cm$^{-1}$ with 
exponent $\alpha'$ and  
(iii) a second, low-frequency multiple time scale regime  in the range 
1- 40 cm$^{-1}$ with a different exponent $\alpha$.
A white noise region ($\alpha=0$), seen at very low frequencies, becomes
more pronounced at higher temperatures and/or higher densities, as discussed
below. The librational peak is seen  in  \sp{u}, \sp{H} and 
\sp{lib}  while the multiple time-scale regions are seen in
all  power spectra except \sp{lib}. At a given state point,
the exact values of the exponents
in the \onebyf\ regimes as well as the crossover frequencies depend
on the specific quantities studied.

Figure 2 shows \sp{u}  at different densities for the
230K isotherm.  At this temperature, SPC/E water has  a diffusional
minimum and maximum at $\rho_{min}$=0.9 g~cm$^{-3}$ and 
$\rho_{max}$=1.1 g~cm$^{-3}$ respectively \cite{nsbs,sslss}. The 
librational region of the \sp{u} curves changes qualitatively 
with density along this isotherm.  It 
is most clearly  demarcated  at $\rho_{min}$=0.9 g~cm$^{-3}$
indicating that decoupling of librational
modes from the hydrogen-bonded network rearrangements is
maximal at this ``ice" density.  The librational
peak moves to lower frequencies with increasing density and, 
for $\rho > \rho_{max}$, becomes
a shoulder. We thus provide direct dynamical evidence 
that the anomalous diffusional regime,
for which $dD/d\rho > 0$, is associated with increasing coupling of the
librational modes with vibrations of the hydrogen-bonded network. This
has been deduced in earlier work based on static structural distributions
\cite{nsbs}.  The density-dependent changes in the \onebyf\ regions
of the power spectra  are quantified by examining the associated exponents. 
Figure 3 shows that
at 230K, the diffusivity, $D$, is strongly correlated
with the low-frequency (1-40 cm$^{-1}$) exponent, $\alpha_u$, and
anticorrelated with the high frequency (60-298 cm$^{-1}$) exponent $\alpha_u'$.

To understand the trends in the exponents of the \onebyf\ regions, 
it should be noted that power spectra
generated using a uniform distribution of time scales
between frequency limits $\lambda_1$ and $\lambda_2$, show
a $1/f$ regime ($\alpha \approx 1$) for $\lambda_1 < f < \lambda_2$, 
a white noise
regime for $f \ll \lambda_1$ ($\alpha=0$) and a Lorentzian tail ($\alpha\approx
2$) for $f\gg\lambda_2$ \cite{em}. Thus $\alpha_u$ and $\alpha_u'$ will
index the shapes of the multiple-time scale regime and the Lorentzian
tail respectively. A separation of two  decades
in time scales between $\lambda_1$ and $\lambda_2$ is adequate 
for observing a \onebyf\ regime with $\alpha$ close to 1.
At $\rho_{min}$,  the librational modes will
be most effectively decoupled from the network vibrations,
the range of available frequencies
contributing to the $1/f$ regime will be the narrowest
and the Lorentzian tail will be  most pronounced. This is 
consistent with our observation  that $\alpha_u'$ is maximum 
and $\alpha_u$ is $\approx 1$ at $\rho_{min}$. In physical terms, it implies
that at 230K and close to $\rho_{min}$, vibrational modes involving three
or less molecules are significantly decoupled from more
delocalised network re-organisations.
We focus on the correlation with the
diffusivity of the exponent $\alpha_u$ associated with the
low-frequency, multiple scale region of the power spectrum.
As librational modes become progressively
more coupled to the network vibrations,
high frequency components are introduced 
resulting in an increase in $\alpha_u$ with density in the
anomalous regime between $\rho_{min}$ and $\rho_{max}$.
 The decrease in $\alpha_u$ with increasing density for $\rho > \rho_{max}$
is due to increasing importance of  steric effects which
force the system to behave more like a simple liquid with $\alpha =0$
\cite{sor92}. 

The behaviour of the librational peak  
in \sp{lib} and \sp{H} at 230K is similar to that seen for \sp{u} 
and the density dependence of the exponents in the \onebyf regions
is  shown in Figure 3.
The $\alpha_H$ and $\alpha_O$ values
are close to 1 in both cases.
The density-dependent variations in  these
quantities, $\alpha_H$ and $\alpha_O$, are smaller in magnitude
than those of $\alpha_u$ though they show similar trends.

Increasing temperature or density results in coupling of librational
modes to the network vibrations and at 300K, the librational band
profile is virtually invariant with density. At 260K, two \onebyf\ regimes
with different slopes can be seen with a low-frequency regime between
1-40 cm$^{-1}$ and a high-frequency Lorentzian tail between 60-200 cm$^{-1}$.
At 300K, only a single \onebyf\ regime can be seen between 4 and 200
cm$^{-1}$ with a crossover to white noise behaviour between
0.1 and 0.5 cm$^{-1}$. Figure 4 illustrates very clearly the correlation
between the density-dependent variations in the diffusivity
and the exponent $\alpha_u$.  The attenuation of the diffusional anomaly 
with temperature and
its disappearance at 300K are  reflected in the behaviour of $\alpha_u$.

We have also examined the \sp{O} and \sp{H} spectra at 260K and 300K.
Compared to \sp{u}, the crossover to white noise
 behaviour occurs at higher frequencies for these quantities;
for example, between 5 to 10 cm$^{-1}$, for \sp{O} and \sp{H} at 300K.
This indicates that local tetrahedral order decorrelates faster than
fluctuations in the tagged molecule potential energy.

This study establishes that the variations in mobility with pressure 
in the region of the diffusional anomaly
are unambiguously mirrored in the exponent of the \onebyf\ region thus allowing
one to monitor the dynamical effects of coupling the localised
librational modes to the hydrogen-bonded network. The range of frequency
over which \onebyf\ behaviour is observed, including 
the frequency of crossover to white noise,  provides
a quantitative assessment of the time and length scales over which
correlations in the network vanish.  Moreover, \onebyf\ behaviour is 
not  equally pronounced for all observables, due to differing
sensitivities to processes operating on different time scales.

The significance of the present results derive from the
clear connections that can be drawn between 
between three distinctive length and time scale
features of bulk water: the local molecular environment, the 
hydrogen bond network reorganisations and the diffusivity which is
a long-time averaged transport
property. This approach provides a possible
route for understanding the connection between structural order and the
static and dynamic anomalies of water \cite{ed01,mc}
and can be extended to other hydrogen-bonded systems or networked liquids.
Our work also suggests that it would be of interest to examine the connection
between \onebyf\ behaviour and spatiotemporal heterogeneity in
supercooled liquids since the distribution of length and time
scales  in the system is expected to widen as one approaches the
kinetic glass transition. In the context of water, such a kinetic glass
transition is predicted for SPC/E water around 190K at 1 g cm$^{-1}$
 \cite{gbss03,gstc};  the experimental value is around 225K \cite{caa83,tbr04}. 
Near this dynamical transition, water behaves in a very similar manner to 
the binary Lennard-Jones glass-formers with no apparent
effect of the hydrogen bond network. Our \onebyf\ analysis shows  that by
230K, two and three-molecule vibrational modes are shifted into the
Lorentzian tail, indicating that these vibrational frequencies are
progressively decoupled from overall network reorganisations.
As the temperature is reduced further, one would expect
the multiple time-scale regime to shift to lower frequencies and
involve larger clusters; the dynamics of these larger clusters
may be insensitive to  directional bonding and local tetrahedral order. 

Our results  have interesting implications
for  experimental  work on water and aqueous solutions, 
specially in the context of hydration of biomolecules
\cite{dcpfs,mcbf}.  
We show that different observables will have different
degrees of sensitivity to the underlying multiple time-scale dynamics
of the hydrogen-bonded network and therefore the choice of spectroscopic
technique is likely to be significant.
For example,  the behaviour of high-frequency librational modes as
 a function of density can be studied by ultrafast spectroscopy
\cite{feltg}. 
Experimental evidence for \onebyf\ behaviour, or the equivalent stretched 
exponential
behaviour of time-correlation functions, exists from inelastic neutron
scattering, Raman and, most recently, optical Kerr effect measurements
 \cite{whycm,gstc,blzc,tbr04}.  The pressure  or density 
dependence of the exponents has not been examined so far and
our work indicates that these will be of considerable interest \cite{tbr04}.

{\bf Acknowledgements} This work was supported by
the Department of Science and Technology (SP/S1/H-16/2000).
AM  thanks  CSIR, New Delhi for the award of a Senior Research Fellowship.

\begin{center}
{\bf Figure Captions}
\end{center}

\begin{enumerate}
\item Power spectra associated with temporal fluctuations in 
different tagged particle quantities at 230K and 0.9 g~cm$^{-3}$.
The different curves represent \sp{u} (thick solid line), \sp{lib} (solid
line), \sp{NN} (dot-dashed line), \sp{O} (dotted line) and \sp{H} (dashed
line).  Arrows show the O-O stretch (200 cm$^{-1}$),
O-O-O bend (50 cm$^{-1}$) and the librational region (400-1000 cm$^{-1}$)
\cite{ph98}.  Systems at a density of 0.75g cm$^{-3}$ are likely
to be inhomogeneous \cite{nsbs}.

\item Power spectra, $S_u(f)$, at different densities (in g cm$^{-3}$)
along the 230K isotherm.

\item Behaviour of  the exponent associated with the \onebyf\ regimes 
as a function of density at 230K.
 Part (a) shows the exponents 
associated with the  high-frequency \onebyf\ regime with
$\alpha_u'$, $\alpha_H'$  and $\alpha_O'$ evaluated over the frequency
ranges 60-298 cm$^{-1}$, 90-298 cm$^{-1}$ and 200-1000 cm$^{-1}$ 
respectively.
 Part (b) shows the exponents 
associated with the  low-frequency \onebyf\ regime
with $\alpha_u$, $\alpha_H$  and $\alpha_O$  evaluated over the
the 1-40 cm$^{-1}$ range. 
Also shown is the diffusivity, $D$,  in units of $2\times 10^{-6}$ cm$^2$/s.

\item Dependence on density, $\rho$, of (a) the diffusivity, $D$, 
and (b) the exponent $\alpha_u$ 
along isotherms at 230K, 260K and 300K.
The exponent $\alpha_u$ is evaluated using the  low-frequency \onebyf\ regime
from 1-40 cm$^{-1}$ at 230K to 280K and over the frequency range 
4-200 cm$^{-1}$ at 300K. 
\end{enumerate}

\includegraphics[width=8.6cm]{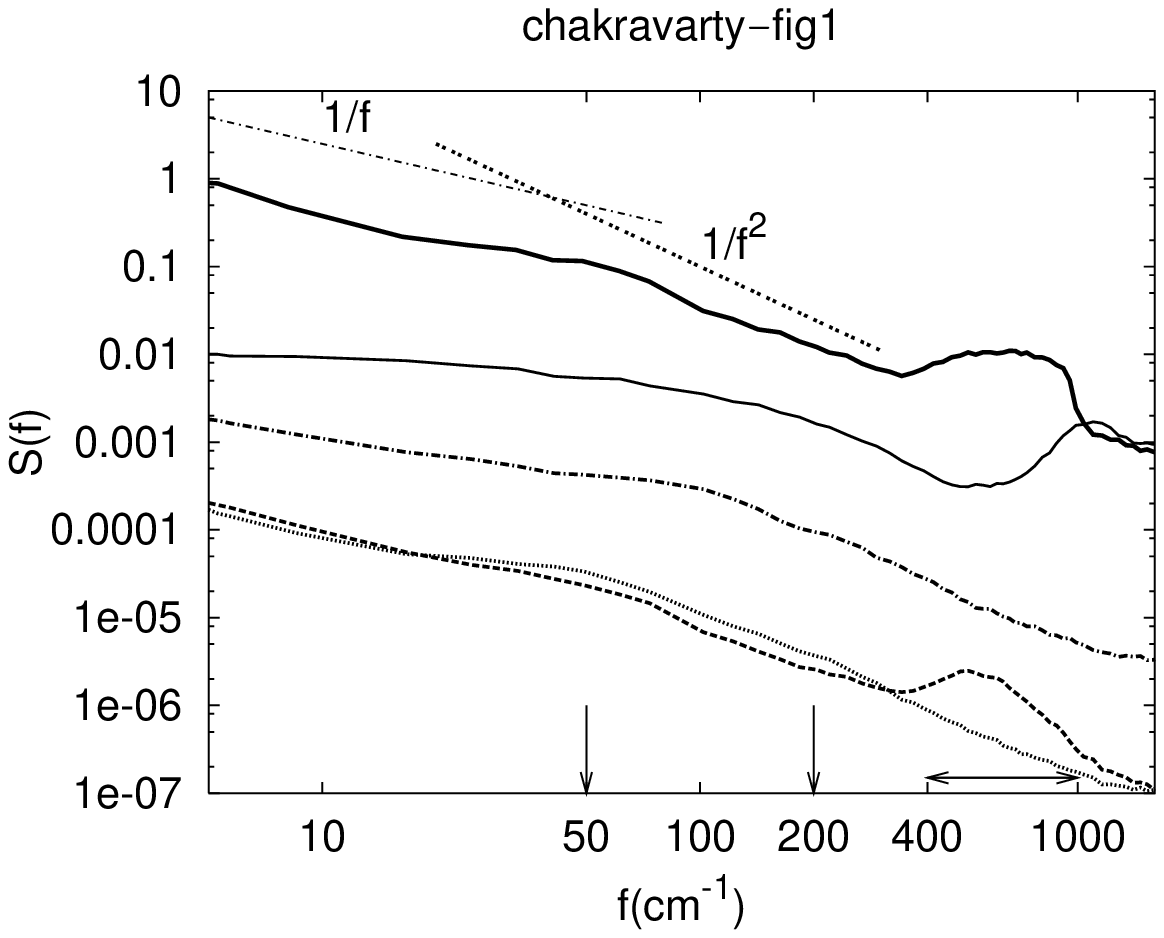}

\includegraphics[width=8.6cm]{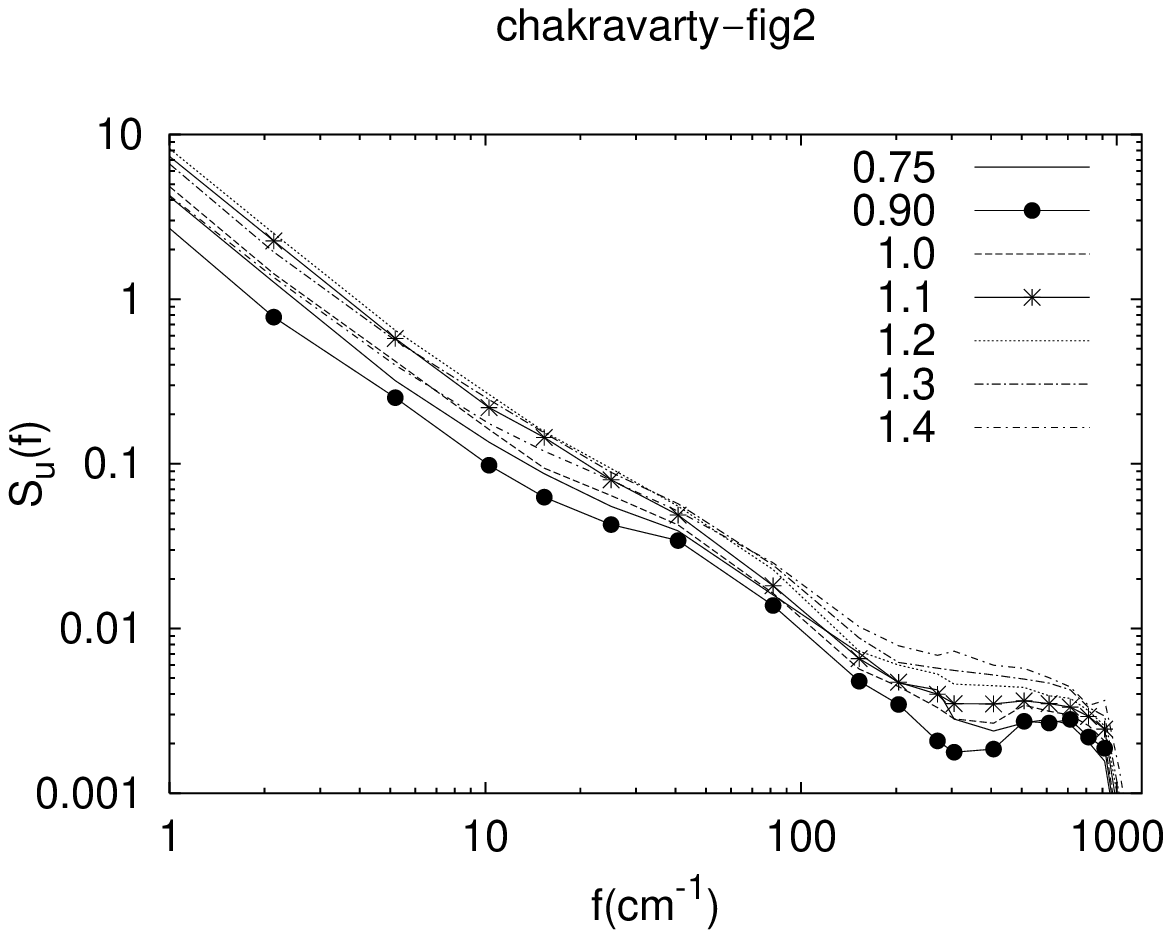}

\includegraphics[width=8.6cm]{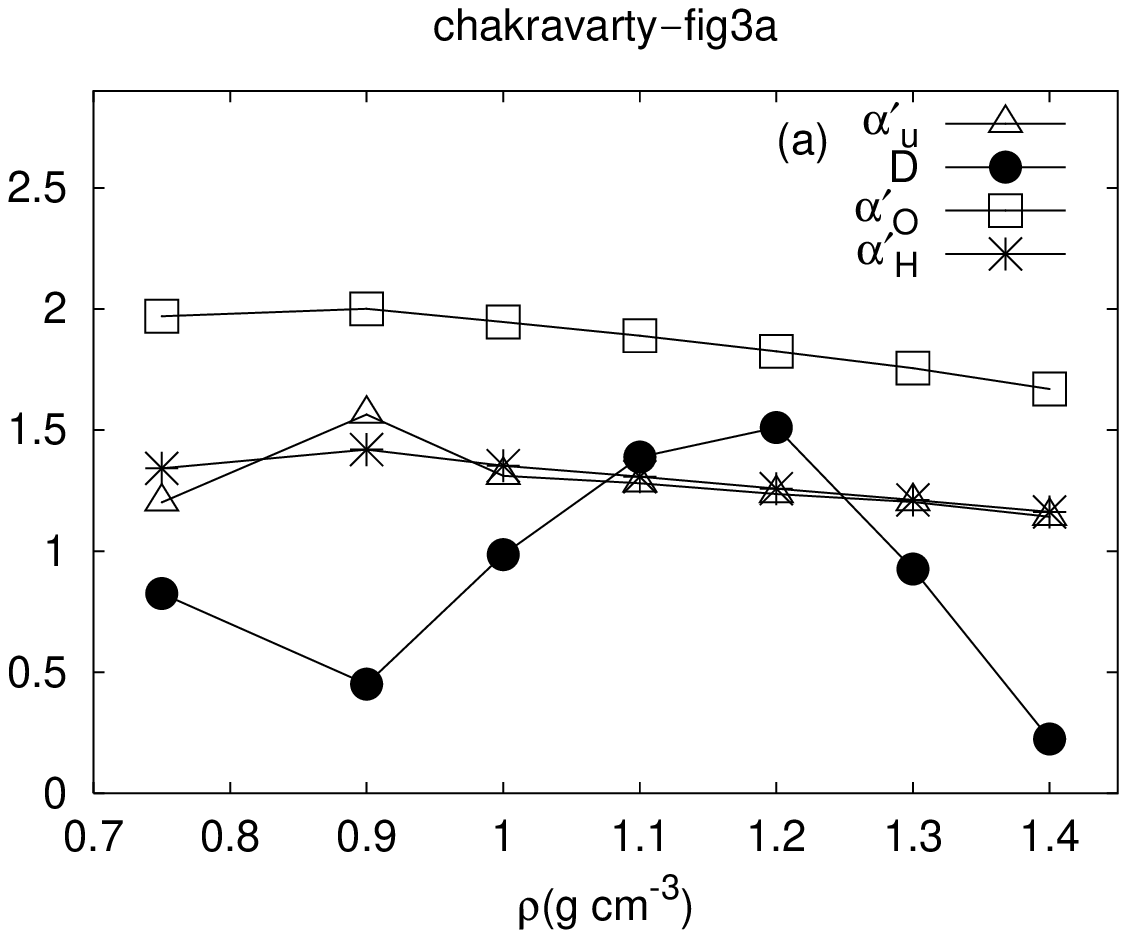}
     
\includegraphics[width=8.6cm]{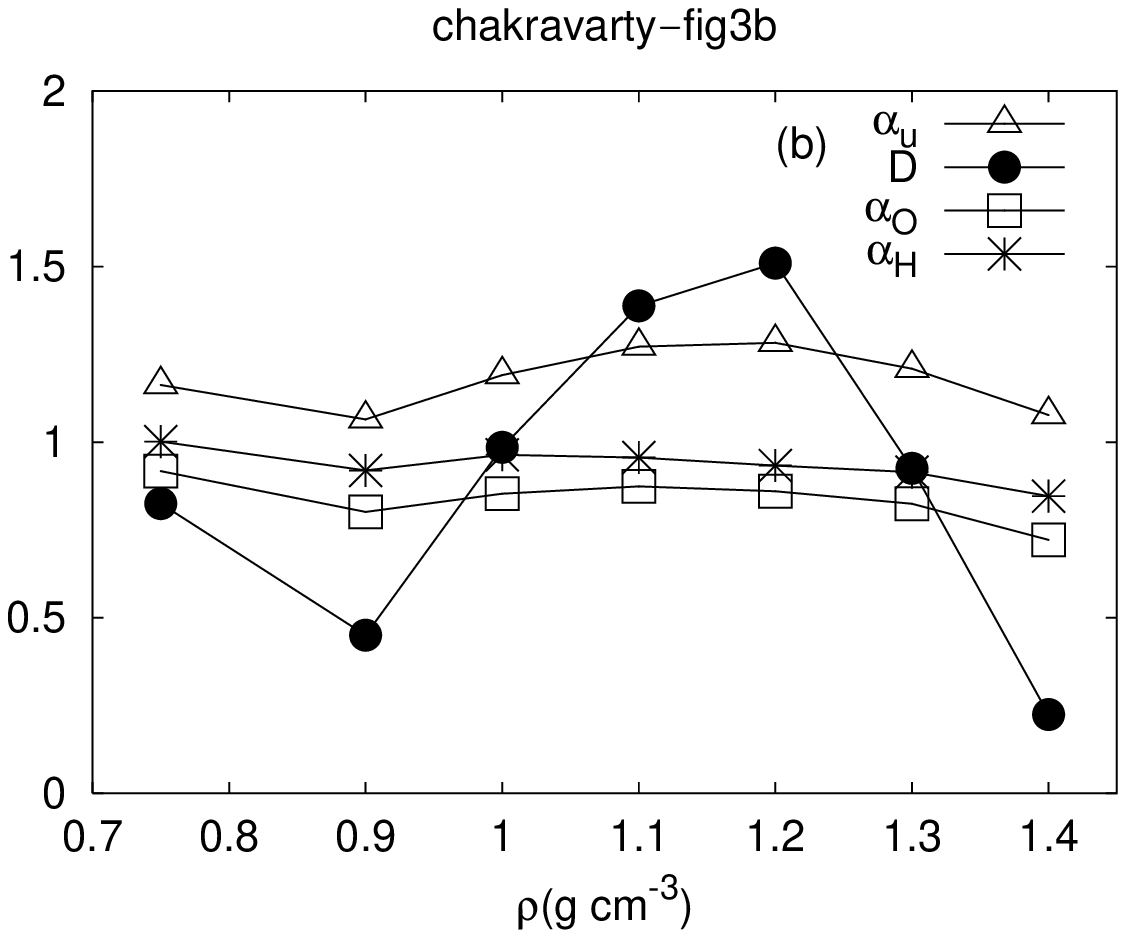}

\includegraphics[width=8.6cm]{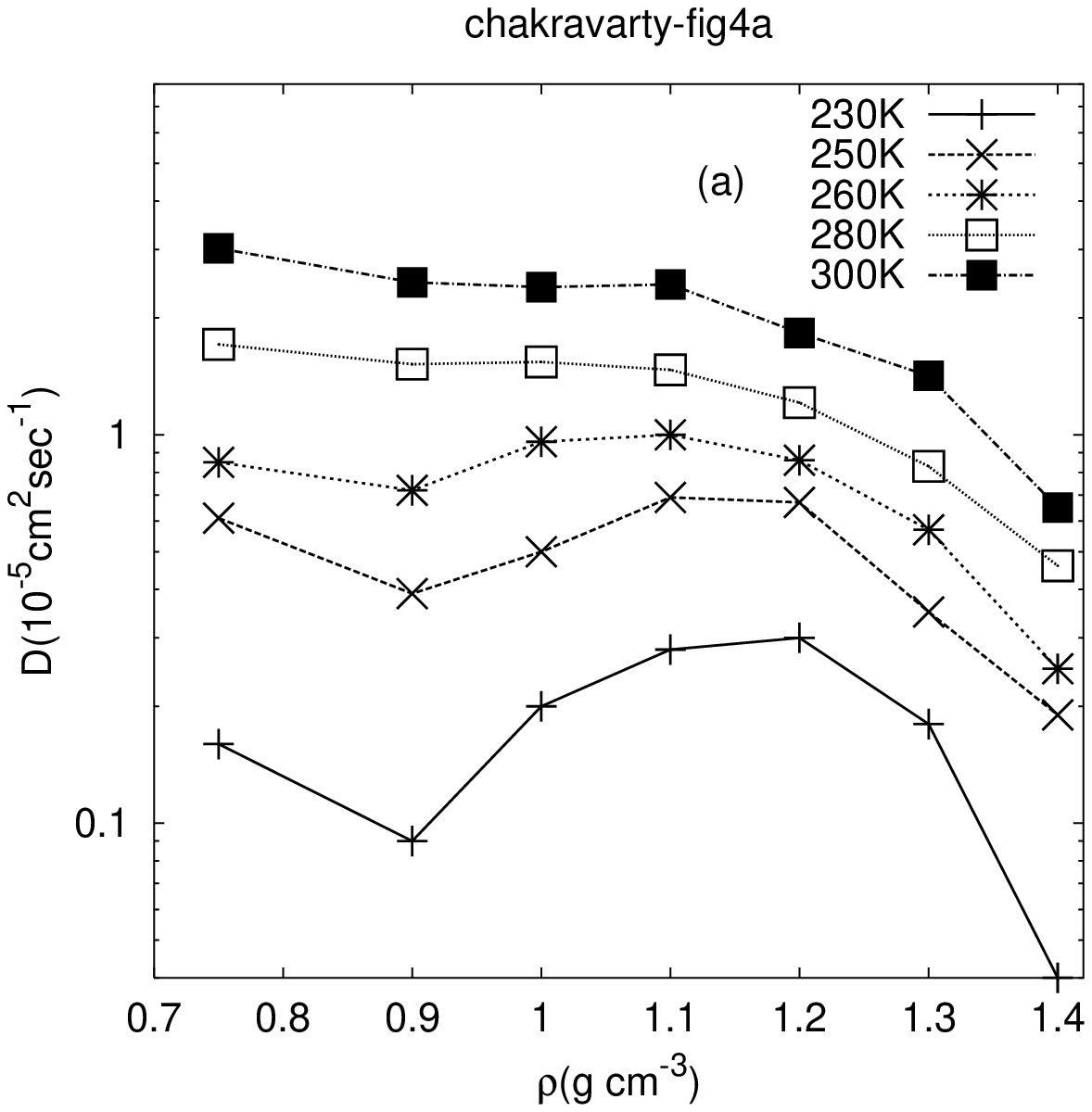}
     
\includegraphics[width=8.6cm]{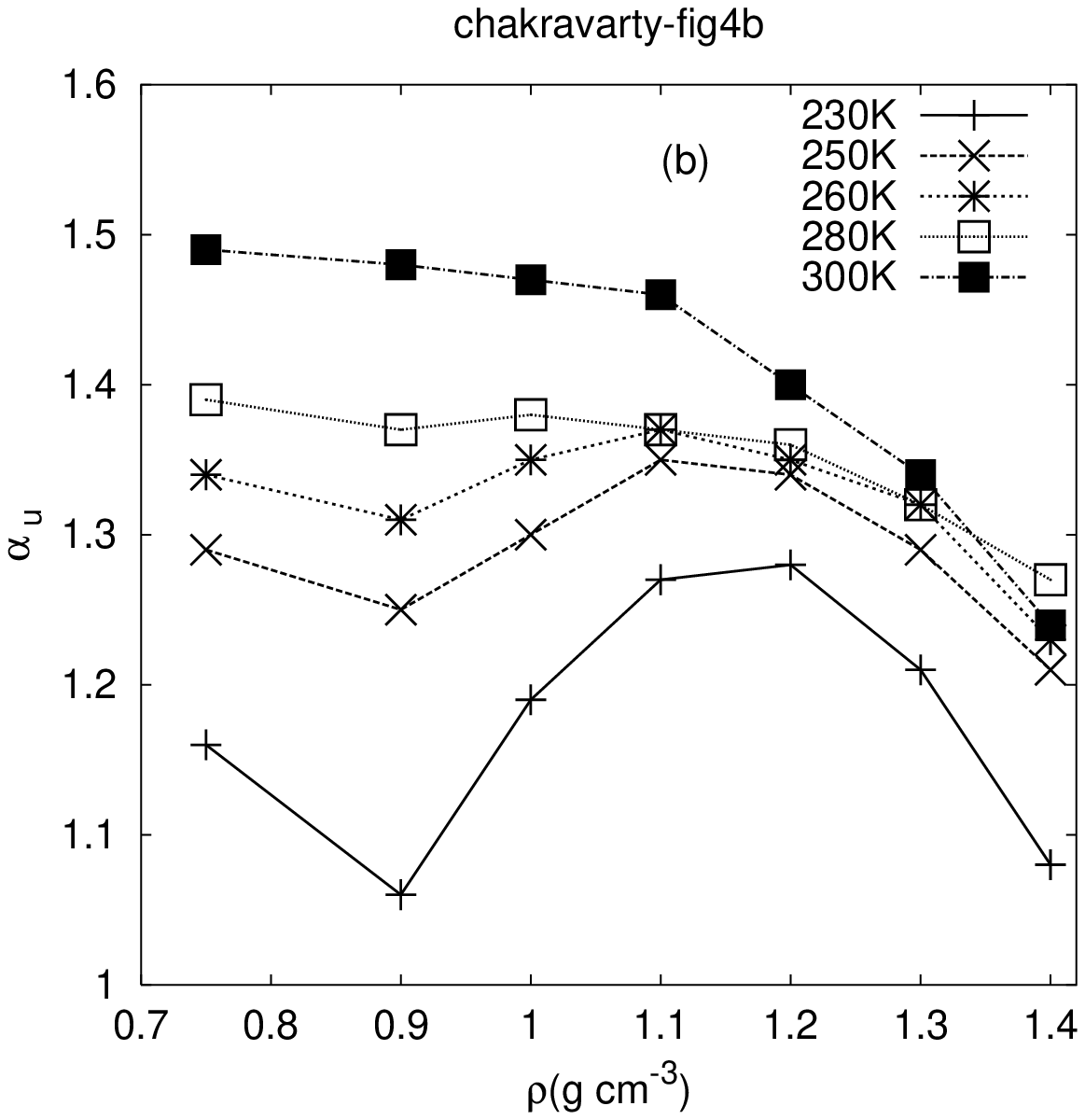}


\begin{thebibliography}{99}
\bibitem{comp} NVT ensemble simulations for 256 SPC/E water
molecules contained in a cubic simulation cell were carried out using the
SHAKE algorithm with a timestep of 1 fs. Electrostatic interactions
were evaluated using Ewald summation. Temperature was maintained
using a Berendsen thermostat with a time constant of 200ps.
\end{thebibliography}
\end{document}